\documentclass[12pt]{article}
\usepackage{amssymb,amsmath,epsfig,mathtools,caption}

\begin{document}

\title{\bf Effects of Thermal Fluctuations on Non-minimal Regular Magnetic Black Hole}
\author{Abdul Jawad$^1$ \thanks{jawadab181@yahoo.com; abduljawad@ciitlahore.edu.pk}
and M.Umair Shahzad$^{1,2}$ \thanks{m.u.shahzad@ucp.edu.pk} \\
\small $^1$Department of Mathematics, COMSATS Institute of Information\\
\small Technology, Lahore-54000, Pakistan.\\
\small $^2$CAMS, UCP Business School, University of Central Punjab,\\
\small  Lahore, Pakistan}

\date{}
\maketitle

\begin{abstract}

We analyze the effects of thermal fluctuations on a regular black
hole (RBH) of non-minimal Einstein-Yang-Mill theory with gauge field
of magnetic Wu-Yang type and a cosmological constant. We consider
the logarithmic corrected entropy in order to analyze the thermal
fluctuations corresponding to non-minimal RBH thermodynamics. In
this scenario, we develop various important thermodynamical
quantities such as entropy, pressure, specific heats, Gibb's free
energy and Helmothz free energy. We investigate first law of
thermodynamics in the presence of logarithmic corrected entropy and
non-minimal RBH. We also discuss the stability of this RBH using
various frameworks such as $\gamma$ factor (comprises of ratio of
heat capacities), phase transition, grand canonical ensemble and
canonical ensemble. It is observed that the non-minimal RBH becomes
more globally and locally stable if we increase the value of
cosmological constant.

\end{abstract}

\section{Introduction}

Black hole (BH) solution is one of the interesting phenomena in
general relativity. Although, its existence is vivid, so it is an
open problem to understand the interior nature of BH in quantitative
detail; the main aspects comes from the fact that a perfect theory
of quantum gravity does not exist \cite{1}. Since the discovery of
Hawking radiation is that, the BH have temperature. Hence, the
concept of BH entropy is no longer a mystery which is proposed by
Bekenstein. Not only that, the work of Hawking proposed the famous
formula of entropy $S=\frac{A}{4}$, where $A$ represents area of
event horizon \cite{2}. BHs have more entropy rather than any other
object of same volume \cite{3,4}. Maximum entropy of BHs is expected
to correct due to quantum fluctuations which leads to the
development of holographic principle \cite{5,6}. As the BH reduces
its size due to Hawking radiation, these fluctuation becomes very
important and are expected to correct the standard relation between
entropy and area \cite{7}.

There are several approaches to evaluate such corrections. Using
non-pertubative quantum general relativity, one can calculate the
density of microstates for asymptotically flat BHs which leads to
the construction of logarithmic correction terms to standard
Bekenstein entropy area relation \cite{8}. One can also use Cardy
formula to generate logarithmic correction terms for all BHs whose
microscopic degrees of freedom are explained by conformal field
theory \cite{9,10}. Ashtekar has obtained such logarithmic
corrections for BTZ BHs by calculating the exact partition function
\cite{8}. These terms could also be generated by the effect of
string theory to the entropy of BH. The analysis of matter fields in
the presence of BH has also generated them for the Bekenstein
entropy area formula \cite{11,12}. In fact, the corrections to the
entropy of dilton BHs are obtained which turns out to be logarithmic
corrections \cite{13}. Rademacher expansion of partition function
can also generate such correction terms \cite{14}. Recently, the
effects of thermal fluctuations on charged ADS BH and modified
Hayward BH has also been investigated \cite{7,15}.

On the other hand, one of the major challenge in general relativity
is the existence of essential singularities (which leads to various
BHs) and it looks like the common property in most of the solutions
of Einstein field equations. Hence, regular black holes (RBHs) have
been constructed to resolve this problem. Since its metric is
regular everywhere, so essential singularities could be avoided in
the solution of Einstein equations of BHs physics \cite{28}. Weak
energy condition is satisfied by these RBHs while some of these
violate the strong energy conditions somewhere in space time
\cite{c17,c18}. Since Penrose cosmic censorship conjecture claims
that singularities predicted by GR occur and they must be discussed
by event horizon \cite{c19,c20}. Hence, Bardeen \cite{29} was the
pioneer who obtained a BH solution without any essential singularity
at origin enclosed by event horizon known as 'Bardeen Black Hole'
which satisfy weak energy conditions. Later on, many authors found
similar solution \cite{30,31,32}. The coupling of general relativity
to non-linear electromagnetic theory has brought to new sets of
charged BHs which came into the range of RBHs solution. Ayon-Beato
and Garcia \cite{30} also found such RBH solution. Hayward \cite{34}
and Berej et. al \cite{33} found different kinds of RBH solutions.
Recently, Leonardo et. al \cite{c22} used many distribution function
in order to obtain charged RBH.

There is an interesting non-minimal theory that couple the
gravitational field to other fields using cross terms of curvature
tensor started to rise long time ago as alternative theories of
gravity. There are five classes of non-minimal field theories
divided accordingly to the types of fields that couple gravitation
to non-minimality, for detail \cite{16a, 16}. These non-minimal
theories construct exact solutions of stars \cite{17,18}, wormholes
\cite{20,21}, BHs \cite{23,24} and regular magnetic BHs \cite{25}
with Wu-Yang anstaz \cite{26,27}. New regular exact spherically
symmetric solutions of a non-minimal Einstein-Yang-Mills theory with
a gauge field of magnetic Wu-Yang and cosmological constant is
presented by Balakin, Lemos and Zayats \cite{16a, 16}. They found
the most interesting solutions of BHs with metric and curvature
invariant are regular everywhere. BH thermodynamics enable us to
study various important thermodynamical quantities of solutions.

One of the most important thermodynamical quantity is thermal
stability of BH. BHs should be stable in dynamical and
thermodynamical frameworks due to their physical nature. The
instability of BHs means whether it may have phase transition or it
is completely unphysical. In this work, we analyze the effects of
thermal fluctuations on RBH of non-minimal Einstein-Yang-Mill theory
with gauge field of magnetic Wu-Yang type and a cosmological
constant. We will use logarithmic correction terms to discuss
various thermodynamical quantities such as pressure, entropy,
specific heats, Gibb's free energy and Helmothz free energy of
non-minimal RBH. The outline of paper is as follows: In section
\textbf{2}, we discuss RBH of non-minimal Einstein-Yang-Mill theory
with gauge field of magnetic Wu-Yang type and a cosmological
constant, furthermore, we will find logarithmic correction terms
which produces various thermodynamical quantities. In section
\textbf{3}, we investigate the stability of non-minimal RBH using
corrected value of specific heat to analyze the phase transition. In
further subsection, we also demonstrate the grand canonical and
canonical ensembles. Conclusion and observations are given in the
last section.

\section{Non-minimal RBH}

New exact regular spherically symmetric solution of non-minimal
Einstein-Yang-Mills theory with magnetic charge of Wu-Yang gauge
field and the cosmological constant is presented by Balakin, Lemos
and Zayats \cite{16a}. Now considering their static spherically
symmetric space-time with line element
\begin{equation}
ds^{2}_e=f(r)dt^{2}-(f(r))^{-1}dr^{2}-r^2({d\theta}^{2}
+\sin\theta{d\phi}^{2}),\label{1}
\end{equation}
where
\begin{equation}\label{2}
f(r)=1+\left(\frac{r^4}{r^4+2Q_m^{2}q}\right)\left(-\frac{2M}{r}+\frac{Q_m^{2}}{r^2}-\frac{\Lambda
r^2}{3}\right),
\end{equation}
is the exact solution to gravitational field equations. This
contains four important parameters such as $\Lambda,~q,~Q_m$ and $M$
which represent the cosmological constant, non-minimal parameter of
the theory, magnetic charge of gauge field Wu-Yang type and mass of
the object, respectively. In this work, we consider $q>0$, $\Lambda
>0$, $\Lambda \leq 0$, $Q_m^{2}>0$ and $M\geq 0$, which have the
following reason. The limiting case $q=0$ gives the magnetic RN
solution with cosmological constant,
\begin{equation}
f(r)= 1+\frac{2M}{r}-\frac{Q_m^{2}}{r^2}-\frac{\Lambda r^2}{3}.
\end{equation}
At $r=0$, we find the curvature singularities. For $q<0$ with finite
positive $r$, we have space-time curvature singularities.

One can obtain no singularities for $q>0$ i.e. $f(r)$ near the
center behaves
\begin{equation}\label{3}
    f(r)= 1+\frac{r^2}{2q^2}-\frac{Mr^3}{Q_m^{2}q}+....
\end{equation}
One can see $f(0)=1$, $f'(0)=0$ and $f''(0)=\frac{1}{q^2}$. Hence
$r=0$ is the minimum of the regular function $f(r)$ which is
independent of cosmological constant and the mass of black hole.
Since $f(0)=1$ and $R(0)=\frac{6}{q}$ shows the curvature scalar is
regular at center. For $q>0$, other curvatures invariants and
quadratic scalar $R_{\mu\nu}R^{\mu\nu}=\frac{9}{q^2}$ are also
finite at center \cite{16}. Thus due to non-minimality of the model,
space-time is truly regular in center.

The metric function (\ref{2}) is described by four parameters with
different units: $\Lambda,~ M,~ Q_m$ and $q$. We can rewrite the
metric function (\ref{2}) in dimensionless form by introducing the
following dimensionless quantities \cite{16a}
\begin{eqnarray}\nonumber
  \gamma_{\Lambda} &=& \sqrt{\frac{3}{|\Lambda|}},~~\gamma_{g}=2M,~~
  \gamma_{q}=\big(2Q_{m}^2 q\big)^{\frac{1}{4}},~~
  \gamma_{Q}=Q_m\\
  \sigma &=& \frac{\gamma_{g}}{ \gamma_{\Lambda}},~~ \eta =
  \Big(\frac{\gamma_{Q}}{\gamma_{\Lambda}}\Big)^2, ~~\zeta =
  \Big(\frac{\gamma_{q}}{\gamma_{\Lambda}}\Big)^4, ~~\rho=
  \Big(\frac{r}{\gamma_{\Lambda}}\Big)^2\label{3a}.
\end{eqnarray}
In terms of these variable the metric function $f(r)$ in (\ref{2})
can be rewritten in $f(\rho)$ as follows
\begin{equation}\label{4a}
    f(\rho)=1+\Big(\frac{\rho^2\big(-\rho^4-\sigma
    \rho+\eta\big)}{\rho^4+\zeta}\Big).
\end{equation}

The most important feature of metric function (\ref{2}) is horizon
radius which depend upon different values of parameters. The horizon
radius of non-minimal RBH could be obtained by considering real
roots of the following equation
\begin{equation}\label{4}
-\frac{\Lambda r_{+}^{6}}{3} + r_{+}^4-2Mr_{+}^3+Q_m^{2}r_{+}^2 +
2Q_m^{2}q=0.
\end{equation}
For $\Lambda > 0$, the non-minimal RBH solution can have three
horizons depending upon different values of parameters, the
cosmological horizon, Cauchy horizon and event horizon. On the other
hand, for $\Lambda \leq 0$, it can have two horizons depending upon
different values of parameters, i.e, Cauchy and event horizons.
There is no cosmological horizon in this case \cite{16a}. Since, we
want to discuss the thermal quantities on outer horizon that is why
we refer $r_+$ as outer horizon in the present work.

One can obtain the mass of the non-minimal RBH in horizon radius and
other parameters as follows
\begin{equation}\label{5}
M=\frac{-\Lambda
r_{+}^6+3Q_m^{2}r_{+}^2+3r_{+}^4+6Q_m^{2}q}{6r_{+}^3},
\end{equation}
which implies that $r_{+}\neq 0$. The entropy of non-minimal RBH
which is related to area of BH horizon is
\begin{equation}\label{6}
    S_0=\pi r_{+}^2,
\end{equation}
and volume is
\begin{equation}\label{7}
   V = \frac{4}{3} \pi r_{+}^3
\end{equation}
The temperature of non-minimal RBH can be written as
\begin{eqnarray}
                T&=&\frac{f'(r)}{4 \pi} \\
                 &=& \frac{-\Lambda r_{+}^9+3Mr_{+}^6+6Q_m^{4}qr_{+}^2-
                 18MQ_m^{2}qr_+^{2}-(6q\Lambda+3)Q_m^{2}r^5}{6 \pi
                 \left(r_{+}^4+2Q^2q\right)^2}\label{8}
              \end{eqnarray}
where the mass $M$  is given in Eq. (\ref{5}). One can examine the
thermodynamics of non-minimal RBH in terms of mass $M$, horizon
radius $r_+$, non-minimal parameter $q$, cosmological constant
$\Lambda$ and magnetic charge $Q_m$. By utilizing Eq.(\ref{5}) in
above expression, the temperature reduces to
\begin{equation}\label{9}
T=\frac{r_{+}^4-\Lambda r_{+}^6-Q_m^{2}r_{+}^2-6Q_m^{2}q}{4\pi
r_{+}\left(r_{+}^4+2Q_m^2q\right)}.
\end{equation}
For modeling of the metric function (\ref{2}) and of the
thermodynamic quantities, the dimensionless parameters (\ref{3a})
are used, but for visibility of the graphs presentation, we use the
following explicit values of different parameters.

For real positive temperature, the following three conditions can be
obtained
\begin{enumerate}
  \item  For $\Lambda = 0.01$ (positive), $Q_m=1$, $q=0.1$, we have
  $1.2\leq r_+ \leq 9.95$.
  \item For $\Lambda = 0$, $Q_m=2$, $q=0.1$, we have
  $r_+ \geq 2.1283$.
  \item For $\Lambda = -0.01$ (negative), $Q_m=3$, $q=0.1$, we have
  $r_+ \geq 2.9713$.
\end{enumerate}
If we variate the values of $\Lambda$ then the range of horizon also
change for real positive temperature. For simplicity, we discuss
above three special cases throughout the paper.

The first law of thermodynamics can be defined as \cite{35,36},
\begin{equation}\label{10}
dM = T dS + V dP + . . . .
\end{equation}
One can easily check the above relation is violated. For obtaining
the thermodynamic quantities which have to satisfy the above
relation, we will use the logarithmic corrections in the following
subsections:

\subsection{Logarithmic correction and thermodynamical relations}

In this section, we discuss the effect of thermal fluctuations on
non-minimal RBH thermodynamics. It is done by using the formalism of
Euclidean quantum gravity, where temporal coordinate is rotated on
complex plane. Hence, one can write the partition function for
non-minimal RBH \cite{7,lc1,lc2,lc3,lc4,lc5}
\begin{equation}\label{lc1}
    Z=\int DgDA \exp(-I),
\end{equation}
where $I\rightarrow iI$ is Euclidean action for this system. One can
relate the statistical mechanical partition function \cite{lc7,lc8}
as
\begin{equation}\label{lc2}
    Z=\int_{0}^{\infty}  DE\eta (E) \exp(-\alpha E),
\end{equation}
where $\alpha=T^{-1}$. We can calculate the density of states by
using
\begin{equation}\label{lc3}
    \eta(E)=\frac{1}{2 \pi i}\int_{\alpha_0-\iota \infty}^{\alpha_0+i \infty}
    d\alpha e^{S(\alpha)},
\end{equation}
where $S=\alpha E+\ln Z.$ This entropy can be obtained around the
equilibrium temperature $\alpha$ by neglecting all thermal
fluctuations which becomes $S_0=\pi r_{+}^2$. However, if thermal
fluctuations are taken into account, then $S(\alpha)$ becomes
\cite{7}
\begin{equation}\label{lc4}
    S=S_0+\frac{1}{2}\big(\alpha-\alpha_0\big)\Big(\frac{\partial^2
    S(\alpha)}{\partial \alpha^2}\Big)_{\alpha=\alpha_0}.
\end{equation}
So, one can write density of states as
\begin{equation}\label{lc5}
    \eta(E)=\frac{1}{2 \pi i}\int_{\alpha_0-i \infty}^{\alpha_0+i \infty}
    d\alpha e^{\frac{1}{2}\big(\alpha-\alpha_0\big)\Big(\frac{\partial^2
    S(\alpha)}{\partial \alpha^2}\Big)_{\alpha=\alpha_0}},
\end{equation}
which leads to
\begin{equation}\label{lc6}
    \eta(E)=\frac{e^{S_0}}{\sqrt{2 \pi}}\Big[\Big(\frac{\partial^2
    S(\alpha)}{\partial
    \alpha^2}\Big)_{\alpha=\alpha_0}\Big]^{\frac{1}{2}}.
\end{equation}
We can write
\begin{equation}\label{lc7}
    S=S_0-\frac{1}{2}\ln \Big[\Big(\frac{\partial^2
    S(\alpha)}{\partial
    \alpha^2}\Big)_{\alpha=\alpha_0}\Big]^{\frac{1}{2}}.
\end{equation}
One can notice that this second derivative of entropy is a
fluctuation squared of energy. It is possible to simplify this
expression by using the relation between the conformal field theory
and the microscopic degrees of freedom of a BH \cite{lc9}. Thus, we
can consider the entropy of the form
$S=m_1\alpha^{n_1}+m_2\alpha^{-n_2}$, where $m_1, m_2, n_1, n_2$ are
all positive constants \cite{lc10}. This has an extremum at
$\alpha_0=\Big(\frac{m_2 n_2}{m_1
n_1}\Big)^\frac{1}{n_1+n_2}=T^{-1}$ and expanding entropy around
this extremum, we can determine \cite{lc11,lc12}
\begin{equation}\label{lc8}
    \Big(\frac{\partial^2
    S(\alpha)}{\partial \alpha^2}\Big)_{\alpha=\alpha_0}=S_0
    \alpha_0^{-2}.
\end{equation}
Thus, the corrected form for the entropy by neglecting higher order
correction terms can be written as
\begin{equation}\label{11}
    S=S_0-\frac{1}{2}\ln S_0 T^2.
\end{equation}
Moreover, the quantum fluctuation in the geometry of BH give rise to
the very important problem of thermal fluctuations in the
thermodynamics of BH. When the size of BH is small and its
temperature is large then it is sufficient to contribute this
correction term. Hence we can avoid the quantum fluctuations for
large BH. It is evident that thermal fluctuation only become
significant for BHs with large temperature and if the size of BH
reduces then its temperature increases. Hence we can conclude that
this corrected terms will only come for sufficiently small BHs which
temperature is large \cite{7}.

Next, we can write the general expression for entropy by neglecting
higher order correction terms
\begin{equation}\label{12}
    S = S_0 - \frac{b}{2} \ln S_0T^2,
\end{equation}
where $b$ is added as constant parameter to handle the logarithmic
correction terms coming from thermal fluctuations. One can recover
the entropy without any correction terms by setting $b=0$. As
mention before, one can take $b\rightarrow 0$, for large BHs which
temperature is very small and one can consider $b\rightarrow 1$, for
small BHs which temperature is sufficiently large. By using Eqs.
(\ref{9}) and (\ref{12}), we can obtain the following corrected
entropy:
\begin{equation}\label{13}
S=\pi
r_+^2-\frac{b}{2}\ln\left(\frac{\left(r_{+}^6\Lambda+Q_m^2r_{+}^2-r_{+}^4+6Q_m^2q\right)^2}{16\pi(r_{+}^4+2Q_m^2q)^2}\right).
\end{equation}
\begin{figure}
  \centering
   \includegraphics[width=8cm]{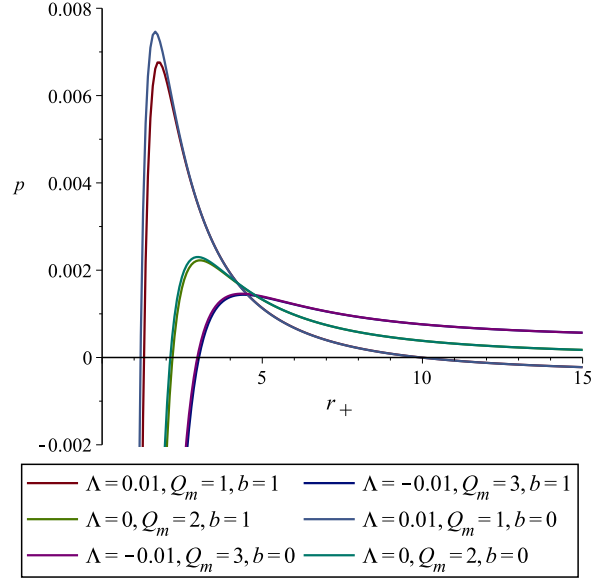}
  \caption{Plot of pressure verses horizon radius.}
\end{figure}
It is suggested that the presence of logarithmic correction causes
the reduction of entropy of BH. We can calculate the pressure using
Eqs. (\ref{7}), (\ref{9}), (\ref{12}), and the following relation:
\begin{equation}\label{14}
 P = T \left(\frac{\partial S}{\partial V}\right)_V.
\end{equation}
which turns out to be
\begin{eqnarray}\label{p14}
    P&=&\frac{1}{8 \pi^2r_{+}^2\big(r_{+}^4+2qQ_{m}^2)^2}\big((b\Lambda+\pi)r_{+}^8+Q_{m}^2r_{+}^4(b q \Lambda
    -4 \pi q-b)\\\nonumber
    &+&2qQ_{m}^4(b-6 \pi q)-\pi r_{+}^{10} \Lambda -\pi Q_{m}^2 r_{+}^6(2q\Lambda+1)-2r_{+}^2qQ_{m}^2(\pi
    Q_{m}^2+8b)).
\end{eqnarray}
In Figure \textbf{1}, we discuss the behavior of pressure for three
cases of $\Lambda$. If we compare $b=0$ and $b=1$, we see that the
pressure decreases due to logarithmic correction. Further, from Fig.
\textbf{1}, we observe that pressure is maximum for positive
cosmological constant but it becomes negative when $r \geq 9.945$,
which is evident in case \textbf{1}. It means that when we increase
the cosmological constant then the pressure also increases but the
range of horizon for positive pressure decreases. {Same behavior
could be observed for temperature. We observe that the pressure is
high for $\Lambda = 0$ as compare to negative cosmological constant.
Hence we conclude that the pressure will decrease due to logarithmic
correction and the lower values of cosmological constant.

Moreover, we can investigate the first law of thermodynamics by
rewriting Eq. (\ref{10}) as follows
\begin{equation}\label{18}
dM-TdS-VdP=0,
\end{equation}
We construct the table for special three cases and find the horizon
at which the first law of thermodynamics is satisfied. We also
compare our results with respect to $b=0$ and $b=1$.
\begin{table}[h]
\begin{center}
\begin{tabular}{|c|c|c|}
\hline
$\Lambda$&\begin{tabular}{@{}c@{}c@{}}Lograthmic \\ correction\\ term\end{tabular}& $r_+$\\
\hline
0.01 & \begin{tabular}{@{}c@{}}$b=0$ \\ $b=1$\end{tabular}&\begin{tabular}{@{}c@{}}$0.4034, 0.9053, 5.53679$ \\ $5.4261, 9.4958, 10.3665$\end{tabular} \\
\hline
0 & \begin{tabular}{@{}c@{}}$b=0$ \\ $b=1$\end{tabular}&\begin{tabular}{@{}c@{}}$0.527, 4.5474$ \\ $2.3\times10^{-5}, 2.4\times10^{-5}, 0.256, 0.329, 1.034, 1.581, 3.343, 4.018$\end{tabular} \\
\hline
-0.01 & \begin{tabular}{@{}c@{}}$b=0$ \\ $b=1$\end{tabular}&\begin{tabular}{@{}c@{}}$0.53269, 5.5951$ \\ $0.2279, 2.5257, 3.6834, 5.45336$\end{tabular} \\
\hline
\end{tabular}
\end{center}
\caption{Location of horizon on first law of thermodynamics
satisfied.}\label{03}
\end{table}
From Table \textbf{1}, we observe that the location of horizon on
which first law of thermodynamics satisfied is more for $b=1$ as
compare to $b=0$. For positive cosmological constant, location of
horizons are equal on $b=0$ and $1$ but for negative cosmological
constant the location of horizons increase for $b=1$ as compare to
$b=0$. We obtain more number of location of horizons on which first
law of thermodynamics is satisfied in the absence of cosmological
constant for $b=1$ as compare to $b=0$. Hence, we can conclude that
logarithmic correction term increases the chance of first law of
thermodynamics to satisfy.

\section{Stability of non-minimal RBH}

In this section, we will analyze the thermodynamical stability of
non-minimal RBH due to the effect of thermal fluctuations. For this
purpose we use the well-known relation
\begin{equation}\label{15}
E = \int T dS.
\end{equation}
We can find the internal energy and observe that it decreases
dramatically due to logarithmic corrections. An important measurable
physical quantity in BH thermodynamics is thermal capacity or heat
capacity. It identifies the amount of heat required to change the
temperature of a BH. The nature of heat capacity (positivity or
negativity) represents the stability or instability of a BH. There
are two different heat capacities associated with a system. $C_p$:
measures the specific heat when the heat is added at constant
pressure and $C_v$: measures the specific heat when the heat is
added to the system by keeping the volume constant. We obtain the
specific heat with constant volume by using the \emph{T} and
\emph{S} as follows
\begin{equation}\label{16}
 C_v = T \left(\frac{\partial S}{\partial T}\right)_V.
\end{equation}
\begin{figure}
  \centering
   \includegraphics[width=8cm]{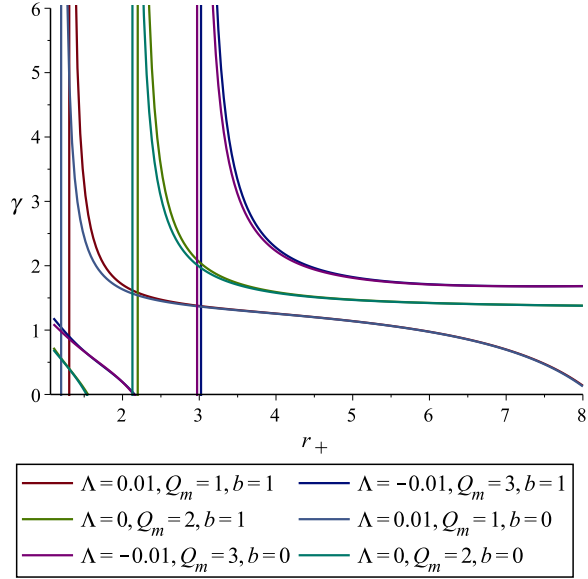}
  \caption{$\gamma$ verses horizon radius.}
\end{figure}
Using Eqs.(\ref{7}), (\ref{9}) and (\ref{12}), we have
\begin{eqnarray}
  C_v &=&\Big(2 \pi r_{+}^{12} \Lambda+4\pi Q_{m}^2qr_{+}^8 \Lambda-2br_{+}^{10}
  \Lambda+2 \pi Q_{m}^2r_{+}^8-12Q_{m}^2 b q r_{+}^6
  \Lambda\\\nonumber
  &-&2\pi r_{+}^{10}+4 \pi Q_{m}^4 q r_{+}^4+8 \pi Q_{m}^2 q r_{+}^6+24 \pi Q_{m}^4 q^2 r_{+}^2
  +2Q_{m}^2 b r_{+}^6-4 Q_{m}^4 b q r_{+}^2\\\nonumber
  &+&32 Q_{m}^2 b q r_{+}^4\Big)\Big(r_{+}^{10}\Lambda+10Q_{m}^2 q r_{+}^6 \Lambda-3 Q_{m}^2 r_{+}^6
  +r_{+}^8+2Q_{m}^4 q r_{+}^2\\\nonumber
   &-& 36 Q_{m}^2 q r_{+}^4-12Q_{m}^4 q^2\Big)^{-1}.\label{c16}
\end{eqnarray}
Moreover, the specific heat at constant pressure can be obtained
using the relation of \emph{E}, \emph{P} and \emph{V} as follows
\begin{equation}\label{17}
 C_p = \left(\frac{\partial(E + PV)}{\partial T}\right)_P.
\end{equation}
Using Eqs.(\ref{7}), (\ref{9}), (\ref{p14}) and (\ref{15}), we have
\begin{eqnarray}
  C_p &=& \Big(64r_{+}^{12} \pi Q_{m}^2 q \Lambda+80 r_{+}^8 \pi Q_{m}^4 q^2
  \Lambda+4 r_{+}^{12}\pi Q_{m}^2-48 r_{+}^{10} Q_{m}^2 b q \Lambda\\\nonumber
  &+&32 r_{+}^8 \pi
  Q_{m}^4 q-48 \pi Q_{m}^2 q r_{+}^10-192 r_{+}^6 Q_{m}^4 b q^2 \Lambda+48 r_{+}^4
  \pi Q_{m}^6 q^2+32 \pi Q_{m}^4 q^2 r_{+}^6\\\nonumber
  &+&192 r_{+}^2 \pi Q_{m}^6 q^3
  +48 r_{+}^6 Q_{m}^4 b q-64 Q_{m}^2 b q r_{+}^8-32 r_{+}^2 Q_{m}^6 b q^2
  +384 Q_{m}^4 b q^2 r_{+}^4\\\nonumber
  &+&12 r_{+}^{16}\pi \Lambda-8 r_{+}^{14} b \Lambda-8 \pi r_{+}^{14}\Big)
  \Big((r_{+}^{10}\Lambda+10Q_{m}^2 q r_{+}^6\Lambda-3Q_{m}^2 r_{+}^6\\\nonumber
  &+&r_{+}^8+2Q_{m}^4 q r_{+}^2-36 Q_{m}^2 q r_{+}^4-12 Q_{m}^4
  q^2)(r_{+}^4+2Q_{m}^2q)\Big)^{-1}.
\end{eqnarray}
The above two specific heat relations can be comprises into a ratio
that is denoted by $\gamma = C_p/C_v$ and its plot is given in Fig.
\textbf{2} for special three cases. We observe that due to
logarithmic correction the value of $\gamma$ increases. We obtain
the maximum value of $\gamma$ for negative cosmological constant and
$\gamma \rightarrow 1.7$ for large horizon. Further, we observe that
$\gamma \rightarrow 1.4$ when $\Lambda = 0$ for large horizon. It is
interesting to note that the value of $\gamma$ decreases drastically
and becomes zero at $r_+ = 9.92$ and the value of $\gamma$ becomes
negative for large horizon. We can say that the value of $\gamma$
shows stable behavior for negative and zero values of cosmological
constant but represents unstable behavior for positive cosmological
constant. Hence, we can conclude that if the value of cosmological
constant decreases then the value of $\gamma$ is higher and exhibits
the more stable behavior but it experiences the unstable behavior
for higher values of cosmological constant.
\begin{figure}
  \centering
   \includegraphics[width=6cm]{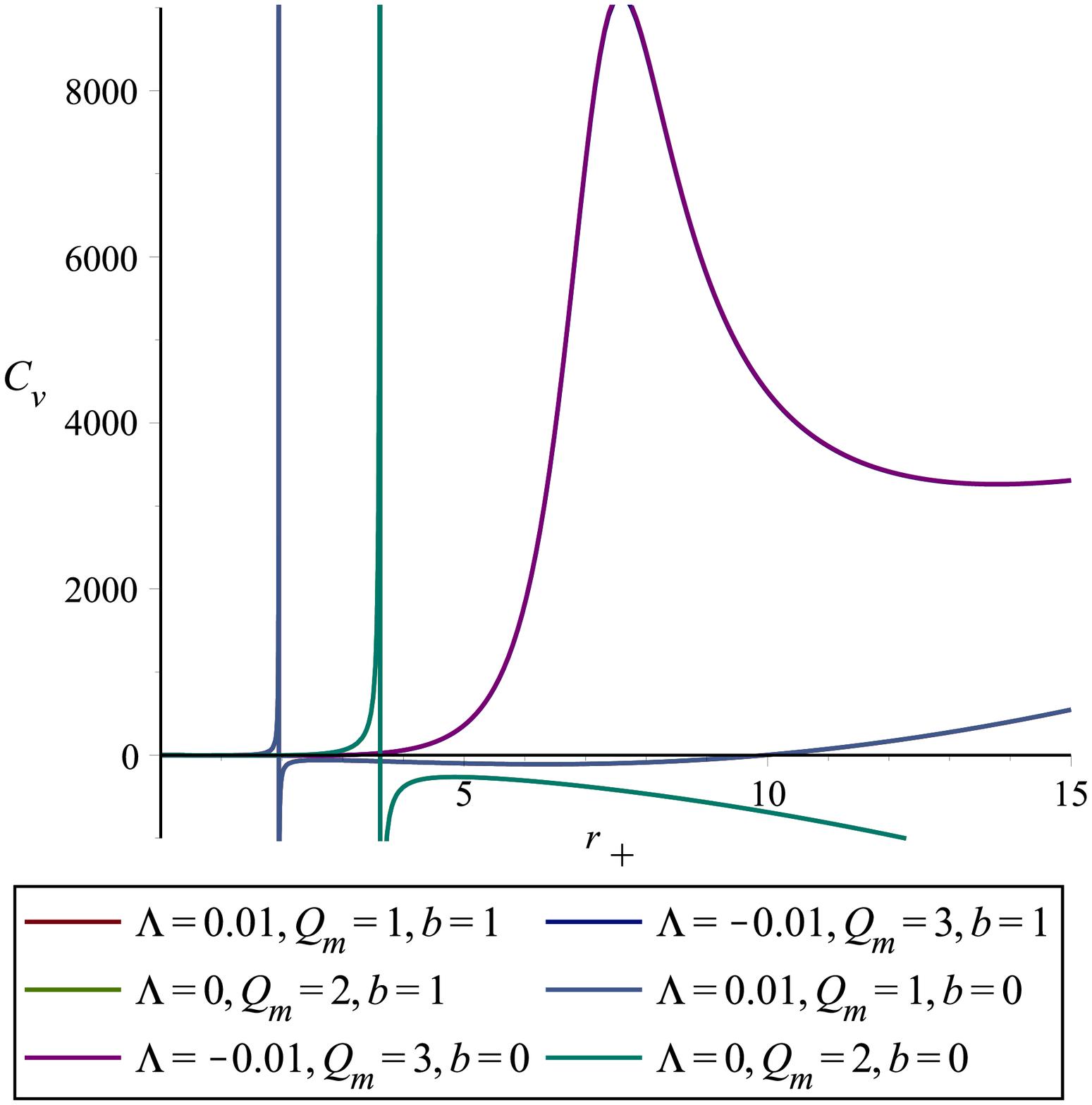}\rlap{\raisebox{2cm}{\includegraphics[width=3cm]{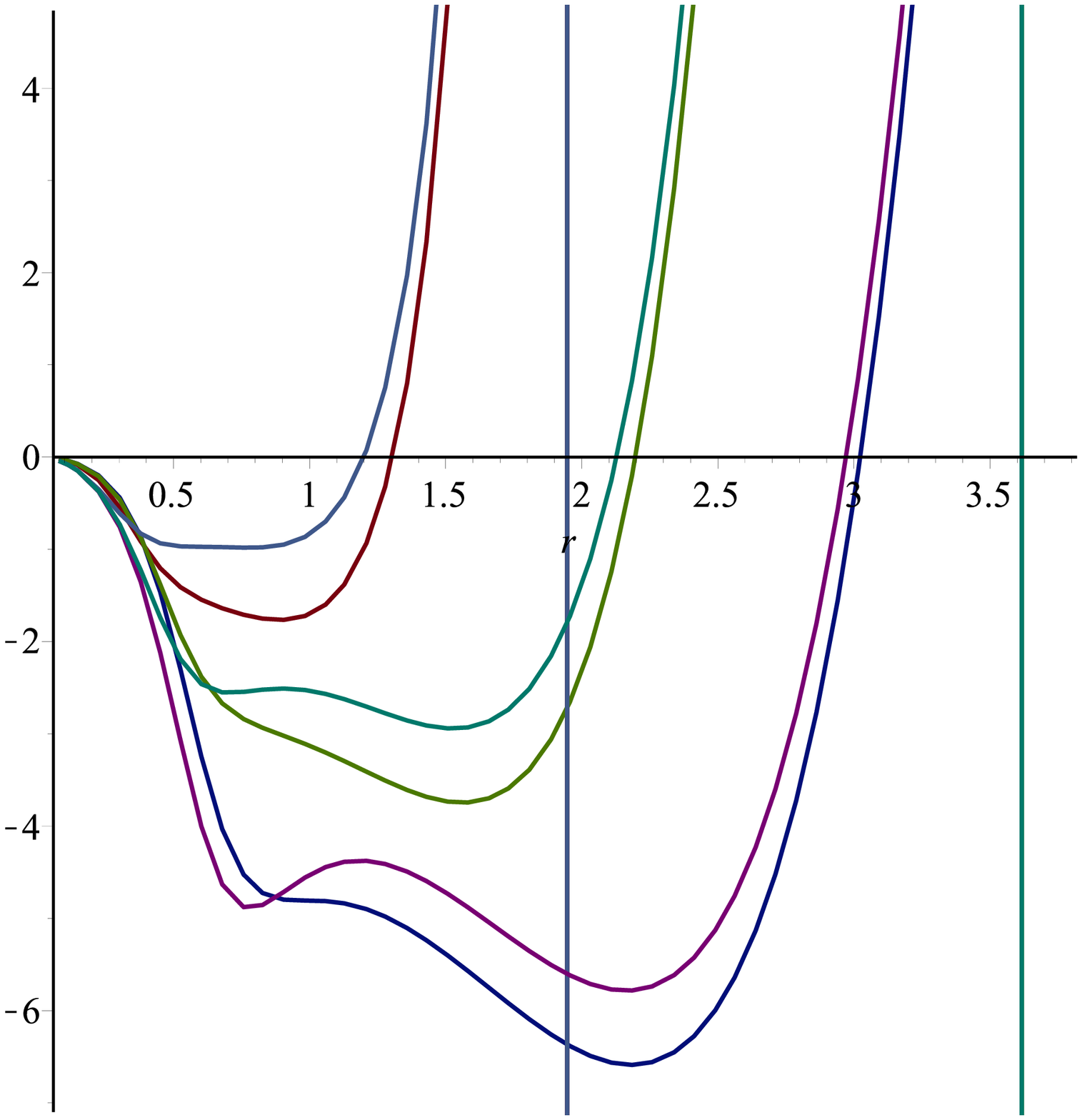}}}\\
  \caption{Specific heat at constant volume verses horizon radius.}
\end{figure}
\subsection{Phase Transition} Another way to find the thermodynamical
stability of BH locally is to investigate the sign of specific heat
given in Eq.(\ref{c16}). The BH is locally stable for $C_v>0$, one
can find the point of phase transition at $C_v=0$ and BH is locally
unstable for $C_v<0$. We can find the range of horizon radius of BH
stability for three specific cases in Fig. \textbf{3}.

We discuss the range of black hole horizon of locally
thermodynamical stability for each case. We observe that when
$\Lambda=0.01$ (positive) the horizon radius for local stability
$1.3047<r_+<1.9$ and $r_+>9.965$, when $\Lambda = 0$ the horizon
radius for local stability is $2.1972<r_+<3.6$ and the horizon
radius is $r_+>3.0232$ for negative cosmological constant. Hence we
can conclude that for negative cosmological constant the range of
horizon radius for local stability of BH is higher as compare to
positive and zero cosmological constant, respectively. Furthermore,
we find the critical point of horizon for phase transition in each
case. We obtain two critical points of phase transition at
$r_+=1.3047$ and $9.965$ for positive cosmological constant, the
critical point for $\Lambda=0$ is $r_+=2.1972$ and the critical
point is $r_+=3.0232$ for negative cosmological constant. We notice
that the phase transition for positive cosmological constant is near
to BH as compare to zero and negative cosmological constant. Hence
we can conclude that if we increase the value of cosmological
constant the phase transition shifted towards the BH and vice versa.

\subsection{Grand Canonical Ensemble}

We may treat BH as a thermodynamical object by considering it as a
grand canonical ensemble system where $\mu = \frac{Q_m}{r_+}$ is a
fix chemical potential. The corresponding temperature and entropy
with logarithmic corrected term are
\begin{eqnarray}\label{19}
T_g = \frac{1}{4 \pi r_+}\left(\frac{r_{+}^2-r_{+}^4 \Lambda- \mu^2
r_{+}^2 - 6 \mu^2 q}{r_+^2+2\mu^2 q}\right),\\\label{20} S=\pi
r_{+}^2-\frac{b}{2} \ln \left(\frac{(r_{+}^6 \Lambda + \mu^2
r_{+}^4+6 \mu^2 q r_{+}^2-r_{+}^4)^2}{16 \pi (r_{+}^4+2 \mu^2 q
r_{+}^2)^2}\right).
\end{eqnarray}
The effect of chemical potential ($\mu$) decreases the temperature.
The free energy in grand canonical ensemble also called Gibbs free
energy can be defined as
\begin{equation}\label{c21}
G = M - T S-\mu Q_m.
\end{equation}
which turns out to be
\begin{eqnarray}\label{21}
    G&=&\frac{-\Lambda
r_{+}^6+3\mu^{2}r_{+}^4+3r_{+}^4+6\mu^{2} r_{+}^2 q}{6r_{+}^3}+
\left(\frac{r_{+}^2-r_{+}^4 \Lambda- \mu^2 r_{+}^2 - 6 \mu^2
q}{r_+^2+2\mu^2 q}\right)\\\nonumber
&\times&\left(\frac{r_{+}}{4}-\frac{b}{8 \pi r_{+}} \ln
\left(\frac{(r_{+}^6 \Lambda + \mu^2 r_{+}^4+6 \mu^2 q
r_{+}^2-r_{+}^4)^2}{16 \pi (r_{+}^4+2 \mu^2 q
r_{+}^2)^2}\right)\right)-\mu^2 r_{+}.
\end{eqnarray}

\begin{figure}
  \centering
   \includegraphics[width=8cm]{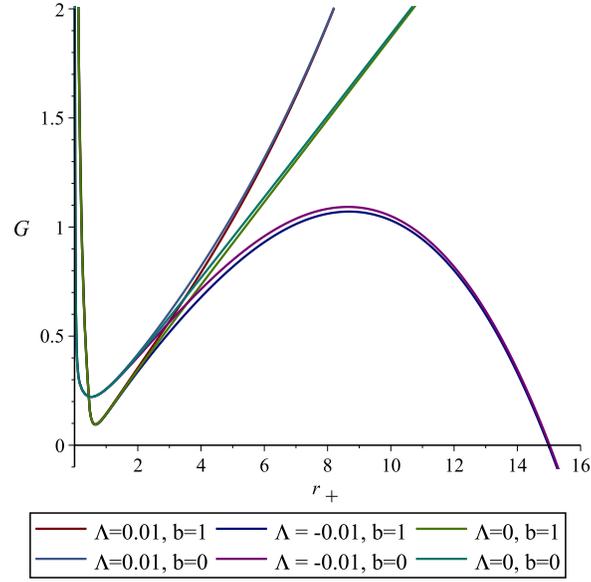}\\
  \caption{Gibb's free energy in terms of horizon radius with $\mu=0.5$.}
\end{figure}
The effect of chemical potential reduces the free energy as we can
see from the last term in Eq.(\ref{21}). Fig. \textbf{4} represents
the behavior of Gibbs free energy for special three cases. We
observe that the Gibbs free energy is minimum at $r_+ \simeq 0.8$
due to the contribution of logarithmic correction term. It means
logarithmic correction term also reduces the Gibbs free energy. From
Figure, we notice that free energy is higher for positive
cosmological constant as compare zero and negative cosmological
constant respectively. It is interesting that free energy becomes
negative at $r_+ \simeq 15$ for negative cosmological constant. It
means free energy is globally thermodynamically unstable for
negative cosmological constant. Hence, we can conclude that if the
value of cosmological constant is higher, then free energy becomes
more globally stable. Moreover, thermodynamical stability does not
only depend on $\Lambda$ and $q$ but also on chemical potential
$\mu$.

\subsection{Canonical ensemble}
\begin{figure}
\centering
\includegraphics[width=8cm]{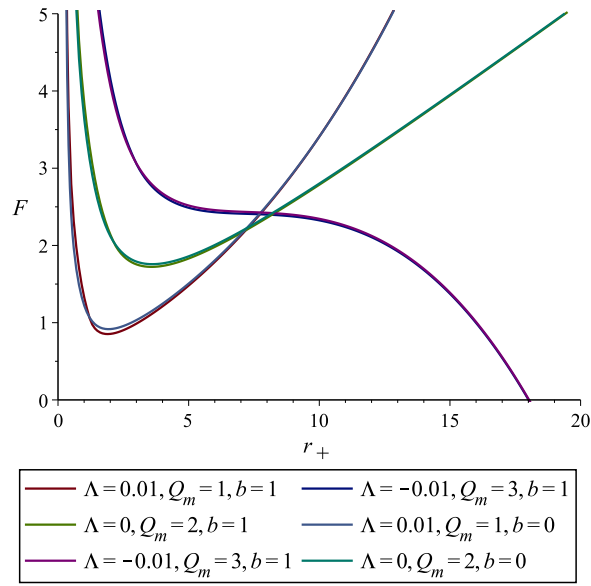}\\
\caption{Helmhotz free energy in terms of horizon radius}
\end{figure}
On the other hand, BH could be considered as a closed system
(canonical ensemble) if the charge transfer is prohibited. The mass
and temperature is given by Eqs.(\ref{5}) and (\ref{9}) respectively
and the corresponding entropy with logarithmic corrected term is
given in Eq. (\ref{13}). The free energy in canonical ensemble is
known as Helmhotz free energy if the charge is fixed, which is
\begin{equation}\label{22}
F = M- TS,
\end{equation}
which turns out to be
\begin{eqnarray}
  F &=& \frac{-\Lambda
r_{+}^6+3Q_m^{2}r_{+}^2+3r_{+}^4+6Q_m^{2}q}{6r_{+}^3}-\frac{r_{+}^4-\Lambda
r_{+}^6-6Q_m^{2}q-6Q_m^{2}q}{4\pi r_{+}\left(r_{+}^4+2Q^2q\right)}
\\\nonumber
&\times&\left(\pi
r_+^2-\frac{b}{2}\ln\left(\frac{\left(r_{+}^6\Lambda+Q_m^2r_{+}^2-r_{+}^4+6Q_m^2q\right)^2}{16\pi(r_{+}^4+2Q_m^2q)^2}\right)\right).
\end{eqnarray}
Fig. \textbf{5} represents the behavior of Helmhotz free energy for
specific values of parameters. We observe that the logarithmic
corrected term reduces the free energy in every case, which is
evident. Initially, the free energy for negative cosmological
constant is high till $r_+=7$ as compare to zero and positive
cosmological constant but for large horizon, free energy is
decreasing further at $r_+ = 18$ it becomes negative. The free
energy is highest for positive cosmological constant as compare to
$\Lambda =0$ and $-0.01$ for large horizon. Hence, we can conclude
that BH is more thermodynamically stable if the value of
cosmological constant increases in large horizon and it becomes
thermodynamically unstable for lower values of cosmological
constant. Moreover, thermodynamical stability do not depend only on
$\Lambda$ and $q$ but also on magnetic charge $Q_m$ rather than
chemical potential.

\section{Concluding Remarks}

In this paper, we have discussed the new exact regular spherically
symmetric solution of non-minimal Einstein Yang-Mill theory with
magnetic charge of Wu-Yang gauge field and the cosmological
constant. We only considered the positive non-minimal parameter $q$
as zero and negative leads to space-time curvature singularities.
After calculating the mass, entropy and temperature, we have
discussed the effect of thermal fluctuation on non-minimal RBH. We
have also utilized the logarithmic correction of entropy and discuss
the behavior of pressure and specific heat. We observed that the
pressure reduces due to the logarithmic correction when decreasing
the value of cosmological constant.

We have also investigated the ratio of specific heat at constant
pressure and volume ($\gamma$), we observe that due to logarithmic
correction the value of $\gamma$ increases. We have also noticed
that the values of $\gamma$ are higher and more stable for negative
values of cosmological constant while it becomes unstable upon
positive values of cosmological constant for large horizon. We
observed that the first law of thermodynamics is satisfied for
non-minimal RBH even in the presence of thermal fluctuations. It is
mentioned here that the logarithmic correction term increases the
chance of first law of thermodynamics to satisfy. We have also
investigated the phase transition for non-minimal RBH and found its
critical points. We observed that the range of horizon radius for
local stability of BH is increased for negative cosmological
constant as compare to positive and zero cosmological constant,
respectively.

We have noticed that if we increase the value of cosmological
constant, the phase transition shifted towards the BH and vice
versa. We have also discussed the free energy in grand canonical
(Gibb's free energy) and canonical (Helmothz free energy) ensembles.
We notice that free energy reduces in the presence of logarithmic
correction. It is concluded that the non-minimal RBH becomes more
stable globally as well as locally if we increase the value of
cosmological constant and vice versa. We have also noticed that the
thermodynamics of non-minimal RBH gets modified because of general
uncertainty principle \cite{38,39}. Such correction terms are
non-trivial which is evident from our results and they may lead to
interesting consequences like existence of BH remnants.

\end{document}